\documentclass[a4paper,11pt]{article}
\usepackage{jinstpub} 
\usepackage{tabularx}
\usepackage{booktabs}

\title{Enhancing Material Screening at Boulby Underground Laboratory with XIA UltraLo-1800 Alpha Particle Detectors}

\author[a,1]{A. Nguyen,
\note{Corresponding author.}}
\author[a,b,2]{S.E.M Ahmed Maouloud,
\note{Corresponding author.}}
\author[a]{X. Liu,}
\author[c]{J.E.Y. Dobson,}
\author[e]{C. Ghag}
\author[d]{L. Le Floch,}
\author[b]{E. Meehan,}
\author[a]{A. St\,J. Murphy,}
\author[b]{S.M. Paling,}
\author[e]{R. Saakyan,}
\author[b]{P.R. Scovell}
\author[b]{and C. Toth}
\affiliation[a]{University of Edinburgh, School of Physics and Astronomy,\\Edinburgh EH9 3FD, UK}
\affiliation[b]{STFC Boulby Underground Laboratory,\\ Boulby, TS13 4UZ, UK}
\affiliation[c]{King’s College London, Department of Physics,\\London WC2R 2LS, UK}
\affiliation[d]{Nantes Université,\\Nantes, France}
\affiliation[e]{University College London (UCL), Department of Physics and Astronomy,\\London WC1E 6BT UK}

\emailAdd{anh.nguyen@ed.ac.uk}

\emailAdd{sahmedm@ed.ac.uk}

\abstract{The Boulby UnderGround Screening (BUGS) facility, located at the Boulby Underground Laboratory, has significantly advanced its material screening capabilities by installing two XIA UltraLo-1800 alpha particle detectors. This study presents a comprehensive evaluation of one of these detectors, operated 1,100 meters underground at the Boulby Underground Laboratory, which provides significant shielding from cosmic radiation and maintains a low ambient radon activity of 2.30 $\pm$ 0.03 Bq/m$^3$. Our evaluation focuses on energy reconstruction accuracy, background radiation rates, and operational stability. The XIA UltraLo-1800 detector demonstrates remarkable stability in energy reconstruction, with less than 0.1 MeV variation over four years. Moreover, the implementation of a graphite-filled PTFE liner in the sample tray resulted in a significant reduction in background radiation levels compared to measurements with the original stainless steel tray, achieving an average activity of 0.15 $\pm$ 0.01 $\alpha$/cm$^2$/khr (kilo-hour, or 1000 hours). Copper sample assays, performed before and after radon exposure, demonstrated the detector's ability to accurately identify and quantify $^{210}$Po contamination. By implementing the robust cleanliness procedures and protocols described in this article, we observed a reduction in $^{210}$Po activity from 0.504 $\pm$ 0.022 mBq to 0.336 $\pm$ 0.013 mBq, highlighting the crucial role of refined cleaning methods in minimizing background for sensitive experiments. Additionally, observations of elevated background activity levels post-high-activity sample measurements illustrate the need for careful management of assay conditions and environment to maintain low background levels. These results highlight the potential of the XIA UltraLo-1800 in enhancing the precision of material assays essential for reducing background interference in rare event experiments.}

\keywords{Detector modelling and simulations I (interaction of radiation with matter, interaction of photons with matter, interaction of hadrons with matter, etc); Noble liquid detectors (scintillation, ionization, double-phase): Low-background radiation detectors}

\arxivnumber{2408.06925} 

\begin{document}
\maketitle
\flushbottom

\section{Introduction}
\label{sec:intro}

The search for rare events, like the hunt for dark matter or the quest to observe neutrinoless double-beta decay, is pushing the boundaries of experimental physics~\cite{cebrian2023review,dolinski2019neutrinoless}. These experiments are incredibly sensitive, needing to detect signals so faint that even the tiniest amount of background radiation can obscure them. The need to find these elusive phenomena has driven significant advances in material screening techniques, from early methods like high-purity germanium (HPGe) spectroscopy to the cutting-edge capabilities of inductively coupled plasma mass spectrometry (ICP-MS), which can pinpoint trace radioactive impurities within the very materials used to build the detectors. This constant refinement of detection methods is critical as scientists strive to reach ever-lower background noise levels. Leading the charge in this effort is the Boulby UnderGround Screening (BUGS) facility, located in the UK's deepest underground laboratory at a depth of 1,100 m (equivalent to 2,840 m of water shielding)~\cite{scovell2023radioassay}.

One of these experiments' biggest challenges is contamination from radon and other background sources. To accurately predict and mitigate these backgrounds, researchers need a complete understanding of the uranium (U) and thorium (Th) decay chains, including alpha-emitting particles deposited on surfaces. U and Th isotopes, with half-lives often exceeding a billion years~\cite{firestone19978th}, are present in trace amounts in virtually every material on Earth – a legacy of the solar system's formation from nucleosynthesis-enriched material~\cite{lodders2003solar}. Radon, a radioactive noble gas, is a key part of these decay chains and is known for its ability to escape from the materials it originates from~\cite{nazaroff1992radon}. Therefore, accurately measuring radon emanation from detector components is crucial for creating realistic simulations of expected background signals in rare-event search experiments.

Among the 36 radon isotopes, ranging in atomic masses from 193 to 228, all are radioactive with the most stable, $^{222}$Rn, constituting part of the $^{238}$U decay chain and exhibiting a half-life of 3.82 days~\cite{firestone19978th}. The next most common isotopes are $^{220}$Rn and $^{219}$Rn, which belong to the $^{232}$Th and $^{235}$U decay chains, respectively~\cite{firestone19978th}. Radon, due to its noble gas characteristics, resists chemical interactions, posing a challenge for its removal~\cite{cox2015allen}. Furthermore, as a monatomic gas, it possesses a long diffusion length in solids~\cite{nazaroff1992radon, szajerski2020numerical}, which, combined with its half-life, results in a resistance to effective sealing. Decay of radon generates electrically charged daughter ions which, due to electrostatic attraction, `plate out' and adhere to nearby surfaces~\cite{stein2018radon}. Subsequent decays of surface deposits may implant themselves to depths of several micrometres, thereby hindering removal efforts~\cite{nazaroff1988radon}. Several radon chain daughter isotopes, notably $^{210}$Po (138 days) and $^{210}$Pb (22 years), have long half-lives against alpha decay. Should these isotopes be in the vicinity of instrument materials with large ($\alpha$, n) cross sections, such as $^{19}$F found in PTFE, then their subsequent alpha decay can produce neutrons within the instrument~\cite{kudryavtsev2020neutron}. These neutrons introduce a problematic background for experiments such as those searching for weakly interacting massive particles, where the signal being searched for is virtually identical to that produced by neutron single scatters.

Over the past decade, the Boulby Underground Laboratory used a number of assay techniques to address the challenge of controlling the background activity during detector construction, including HPGe spectroscopy~\cite{akerib2020lux}, ICP-MS~\cite{dobson2018ultra}, and, more recently, radon emanation detector system. In 2018, two XIA UltraLo-1800 alpha particle detectors were integrated to extend the BUGS facility's material screening capabilities to surface and bulk alpha emissions. This study, therefore, focuses on the following:

\begin{enumerate}
    \item Characterising the XIA UltraLo-1800 detector's performance, focusing on energy reconstruction and background radioactivity levels.
    \item Analysing the detector's stability.
    \item Demonstrating the detector's assay capabilities by assessing the radioactivity of a copper sample before and after exposure to atmospheric radon.
\end{enumerate}

\section{The Boulby XIA UltraLo-1800}

The XIA UltraLo-1800, manufactured by XIA LLC~\cite{xiallc}, is a state-of-the-art alpha particle detector that employs a dual-channel pulse-shape-analysis technique. This method distinguishes alpha particles originating from a tested sample and those emitted by the detector's components.

\begin{figure}[htbp]
\centering
\includegraphics[width=.48\textwidth]{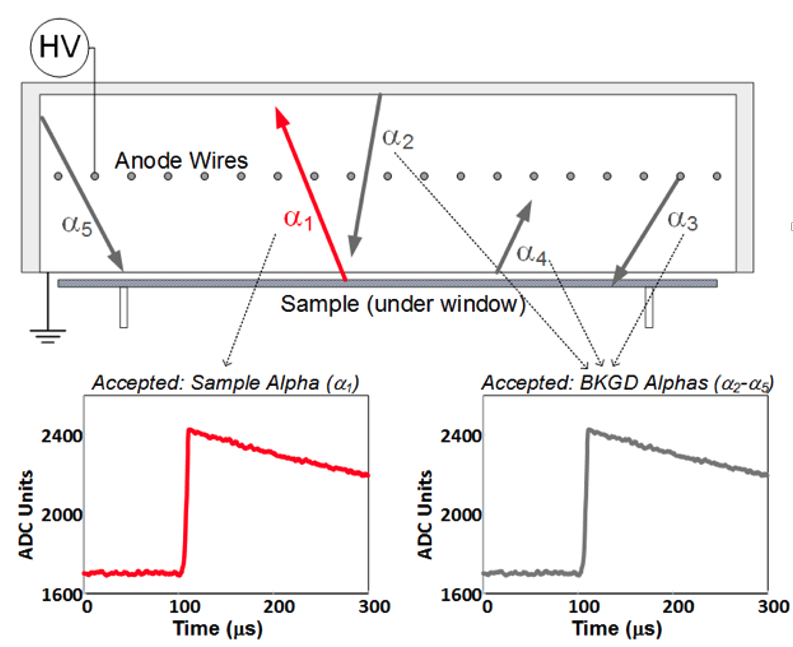}
\qquad
\includegraphics[width=.46\textwidth]{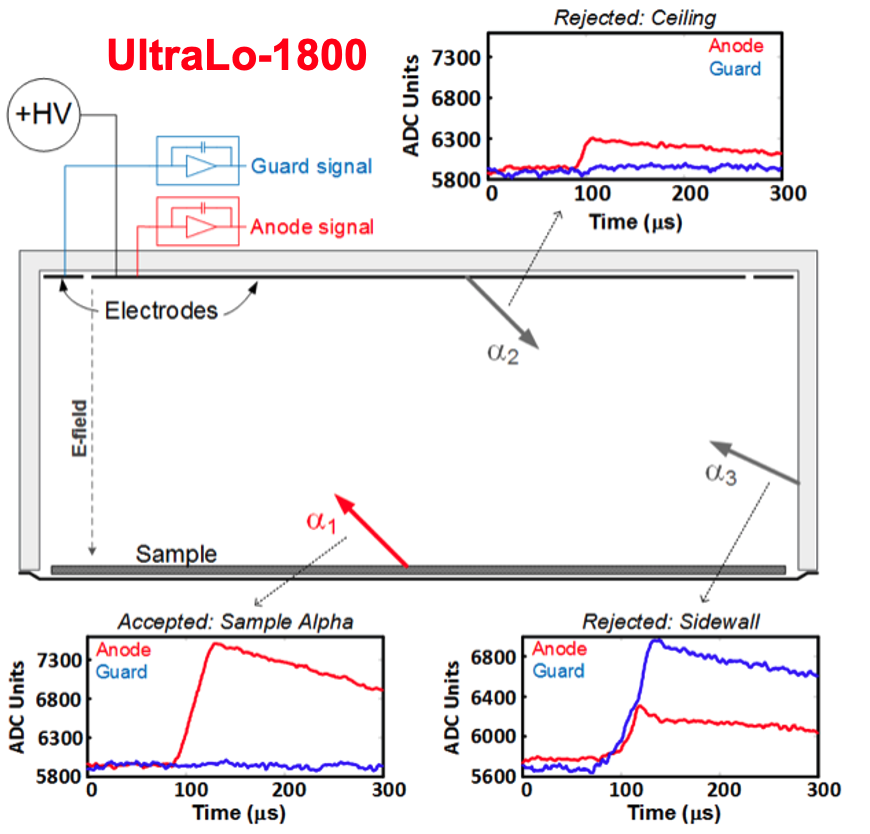}
\caption{Contrast between a conventional proportional counter (left) and the XIA UltraLo-1800 (right) in terms of signal and background detection, reproduced from~\cite{XIA_usersmanual}.\label{fig:xia_schematic}}
\end{figure}

Unlike traditional proportional counters, the UltraLo-1800 operates at relatively lower voltages, which induces ionisation-mode signals, allowing for a more nuanced detection and differentiation of alpha particles~\cite{ange2023characterization}. The dual-channel method uses signal readouts from both an anode located above the sample tray and a guard rail positioned on the sides, as illustrated in Figure~\ref{fig:xia_schematic}. The areas of signal pulses correlate with tracking length, and by comparing anode and guard signals, emissions not derived from the sample can be effectively eliminated. Consequently, these devices have exhibited background rates significantly lower (by a factor of 50 or more) than those observed with conventional proportional counters~\cite{warburton2004,nakib2013,McNally2014, xmass2018}.

Two UltraLo-1800 detectors were installed within a class 1000 cleanroom in the BUGS facility.
Boil-off from a 240~l Dewar, equipped with a `gas use conversion kit' for the regulated flow of gaseous argon, was used to supply the required 99.9995\% purity 
argon gas to the active volume of the UltraLo-1800 counter.

\subsection{Detector Calibration and Energy Reconstruction}

The Boulby UltraLo-1800 was routinely calibrated using a custom-made $^{210}$Po alpha-particle source, 1 inch in diameter, provided by Spectrum Techniques~\cite{st_source}. $^{210}$Po decays by emitting a 5.304~MeV alpha particle with a branching ratio of essentially 100\%~\cite{wang2021ame}.
To better understand the detector's response to various materials and surface contamination, we constructed comprehensive simulations using the GEANT4-v10.5.1 toolkit~\cite{G4}; G4EmStandardPhysics, PhysListEmPenelope, and PhysListEmLivermore were used to model electromagnetic processes, while the G4HadronElasticPhysicsHP was used to model hadronic interactions \cite{ParticleDataGroup:2018ovx, 10.1007/978-3-642-58113-7_227}.
The geometry of the XIA UltraLo-1800 detector was implemented in the simulation, including all physical aspects, such as the active volume, anode, and guard rail, as well as its immediate surroundings. The $^{210}$Po source was modelled as an isotropic source positioned at the centre of the sample tray.

\begin{figure}[htbp]
\centering
\includegraphics[width=.47\textwidth]{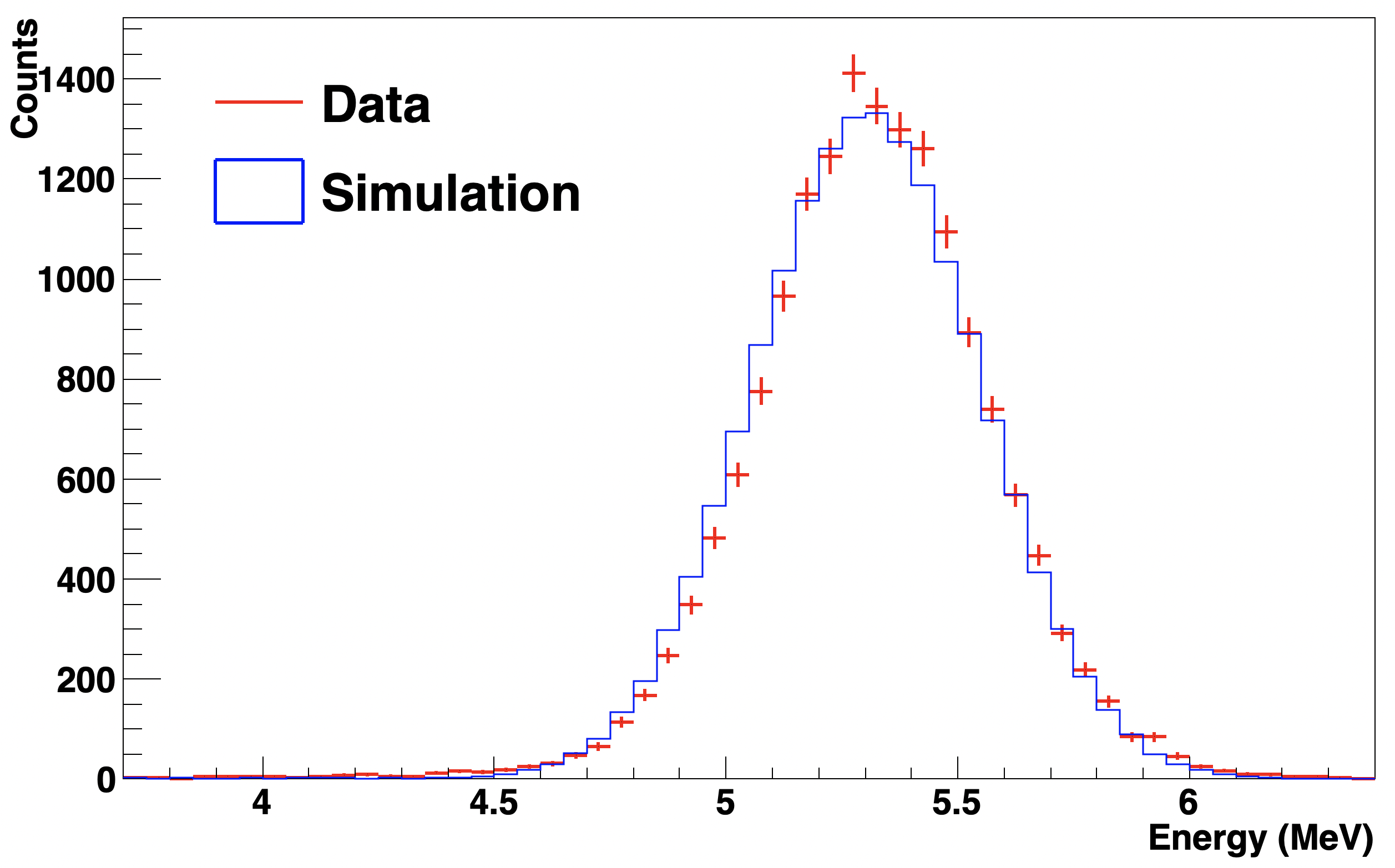} 
\qquad
\includegraphics[width=.47\textwidth]{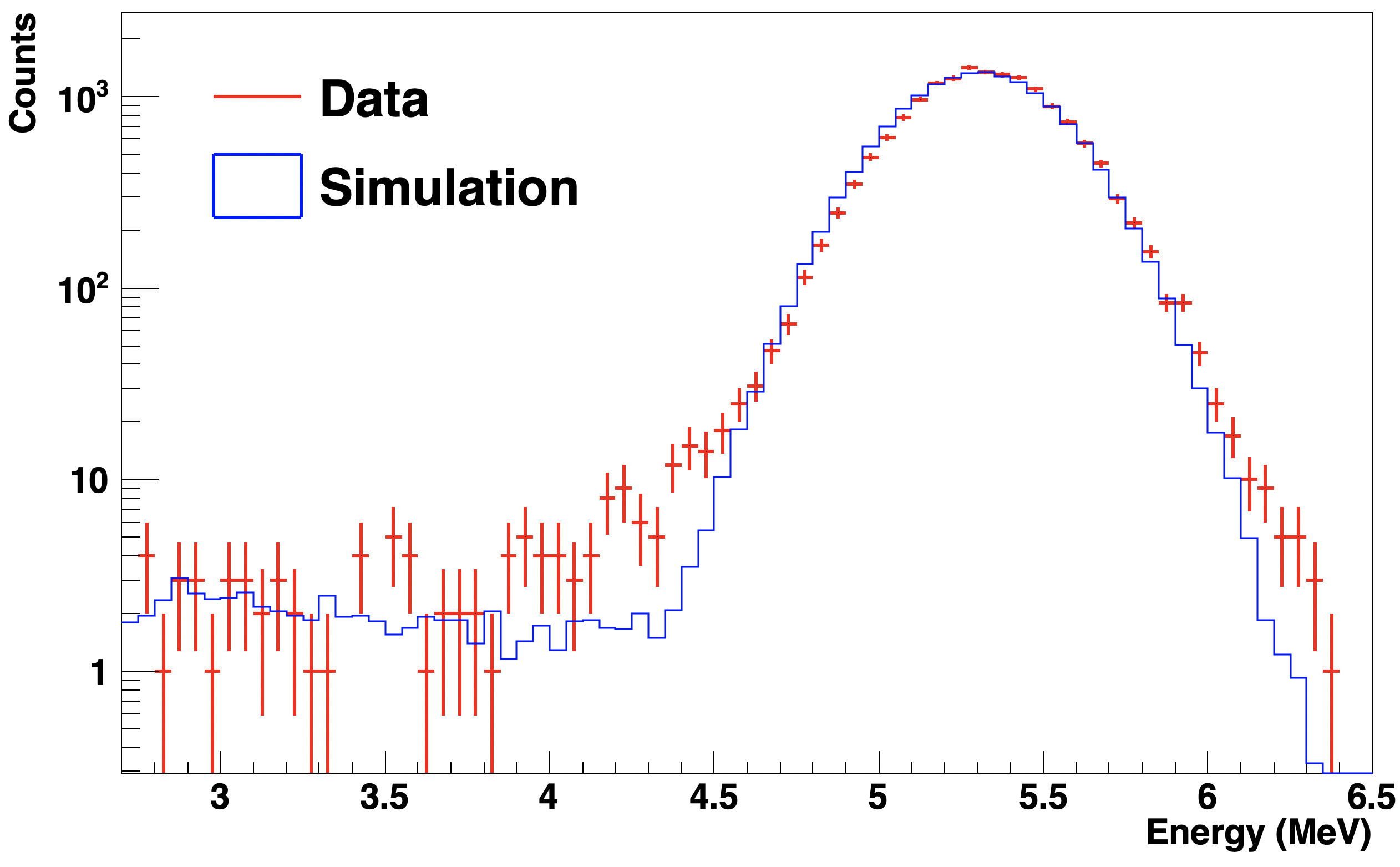}
\caption{Calibration spectra from a $^{210}$Po source on the UltraLo-1800 detector. Data (red) and simulation (blue) are illustrated with linear (left) and logarithmic (right) scales.\label{fig:calibration}}
\end{figure}

Figure~\ref{fig:calibration} displays example calibration data alongside simulated results. The centroid of reconstructed alpha particle energies demonstrated remarkable stability, changing by less than 0.1~MeV over three years. The energy resolution was similarly stable, with a mean value of 0.57~MeV FWHM consistently throughout the measurement period. These findings underscore the robustness of our calibration approach. Our calibration approach, using a $^{210}$Po source, demonstrates good energy reconstruction stability over several years. We extended a previous study by Southern Methodist University (SMU)~\cite{ange2023characterization}, which employed a similar methodology with a $^{230}$Th source, by incorporating an analysis of the low-energy tail observed in the spectral data. This approach ensures a more accurate determination of the detector's efficiency, particularly at lower energy thresholds. In the following section, we will present a detailed analysis of the Boulby UltraLo-1800's long-term stability and the comprehensive nature of our measurements, which address this potential limitation.

\subsection{Background Measurements}

\begin{figure}[htbp]
\centering
\includegraphics[width=.47\textwidth]{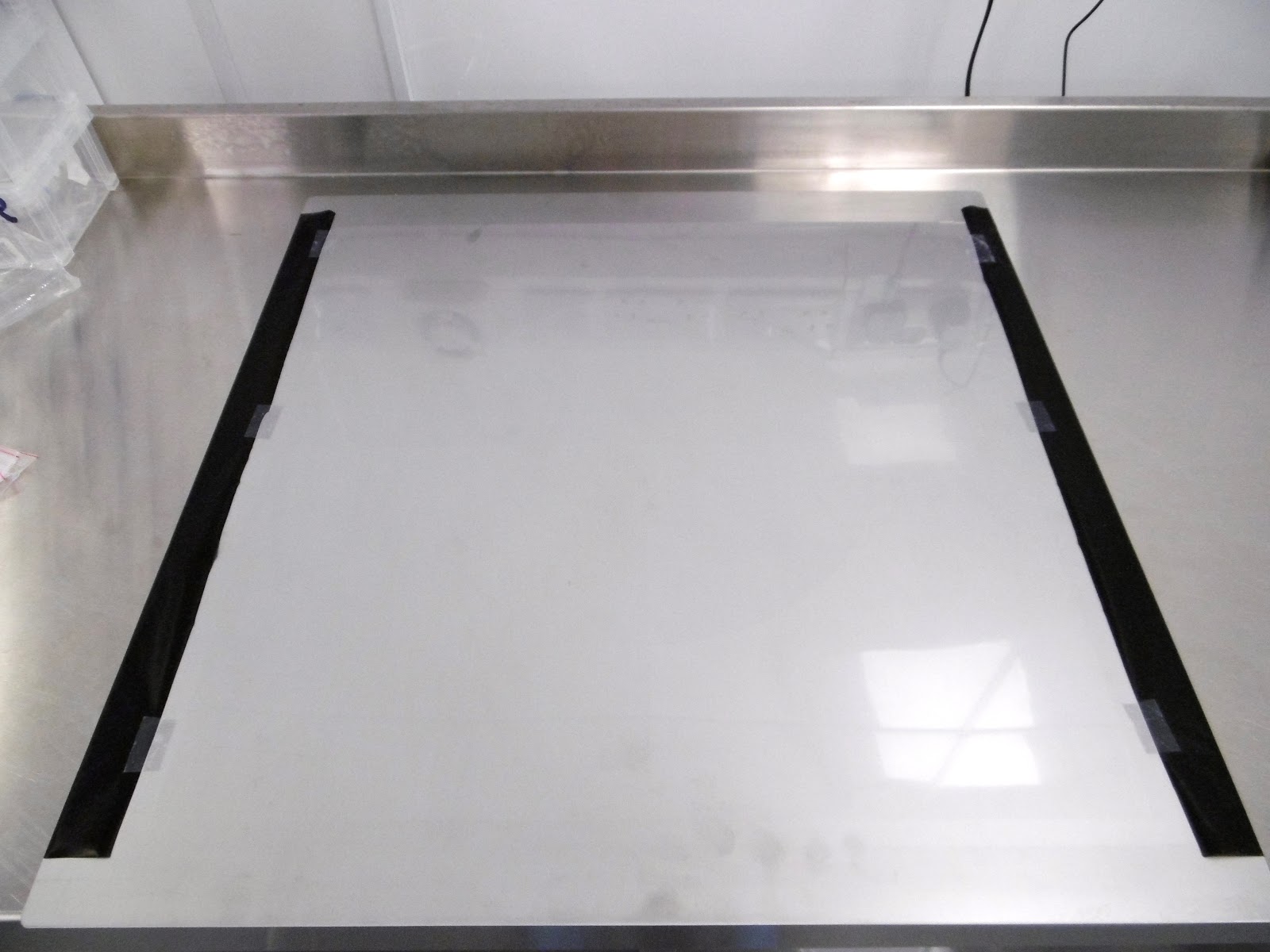}
\qquad
\includegraphics[width=.47\textwidth]{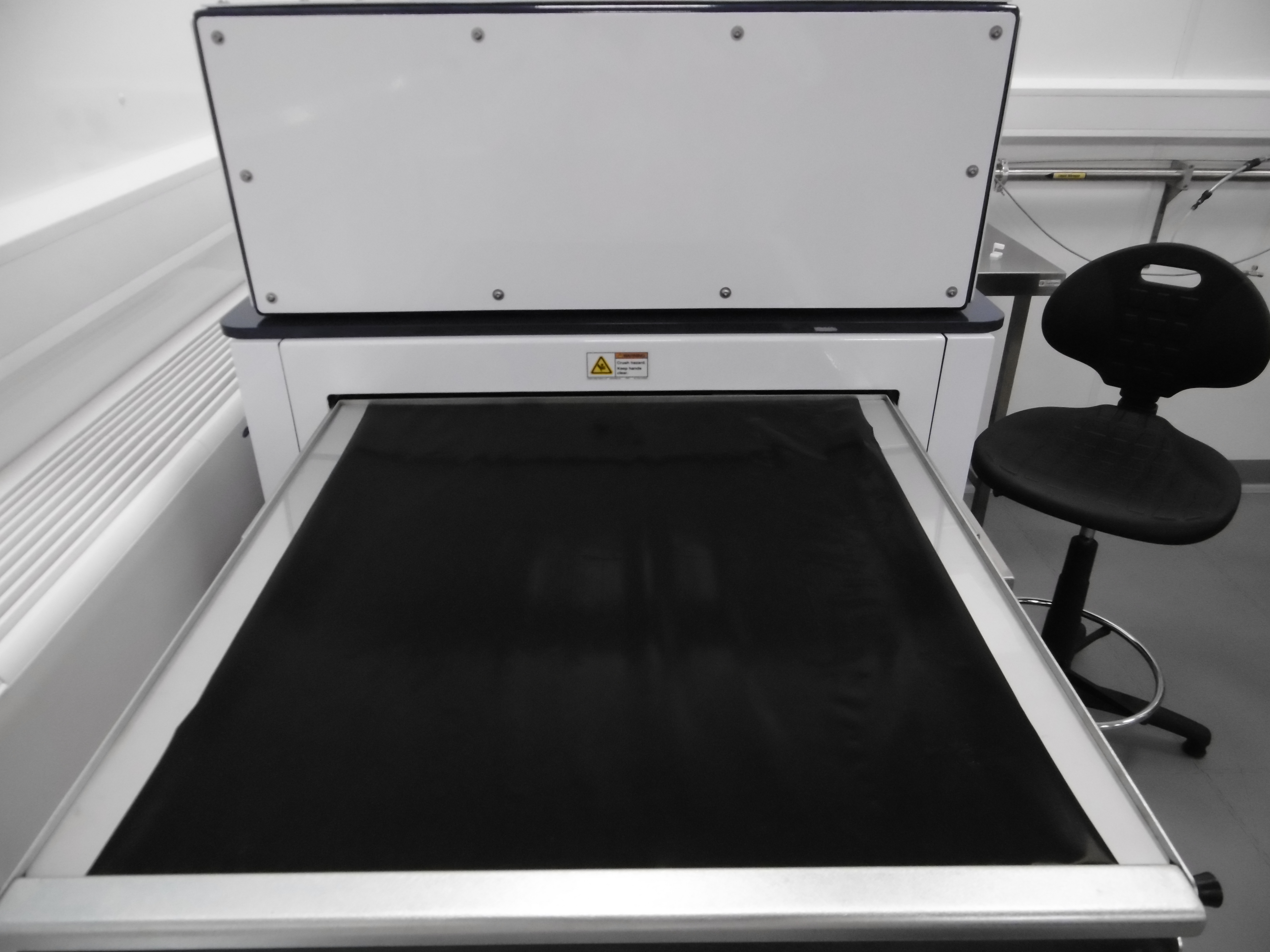}
\caption{Photographs of the UltraLo-1800 stainless steel sample tray (left) only and with the PTFE-graphite liner (right) installed.\label{photo}}
\end{figure}

\begin{figure}[htbp]
\centering
\includegraphics[width=.95\textwidth]{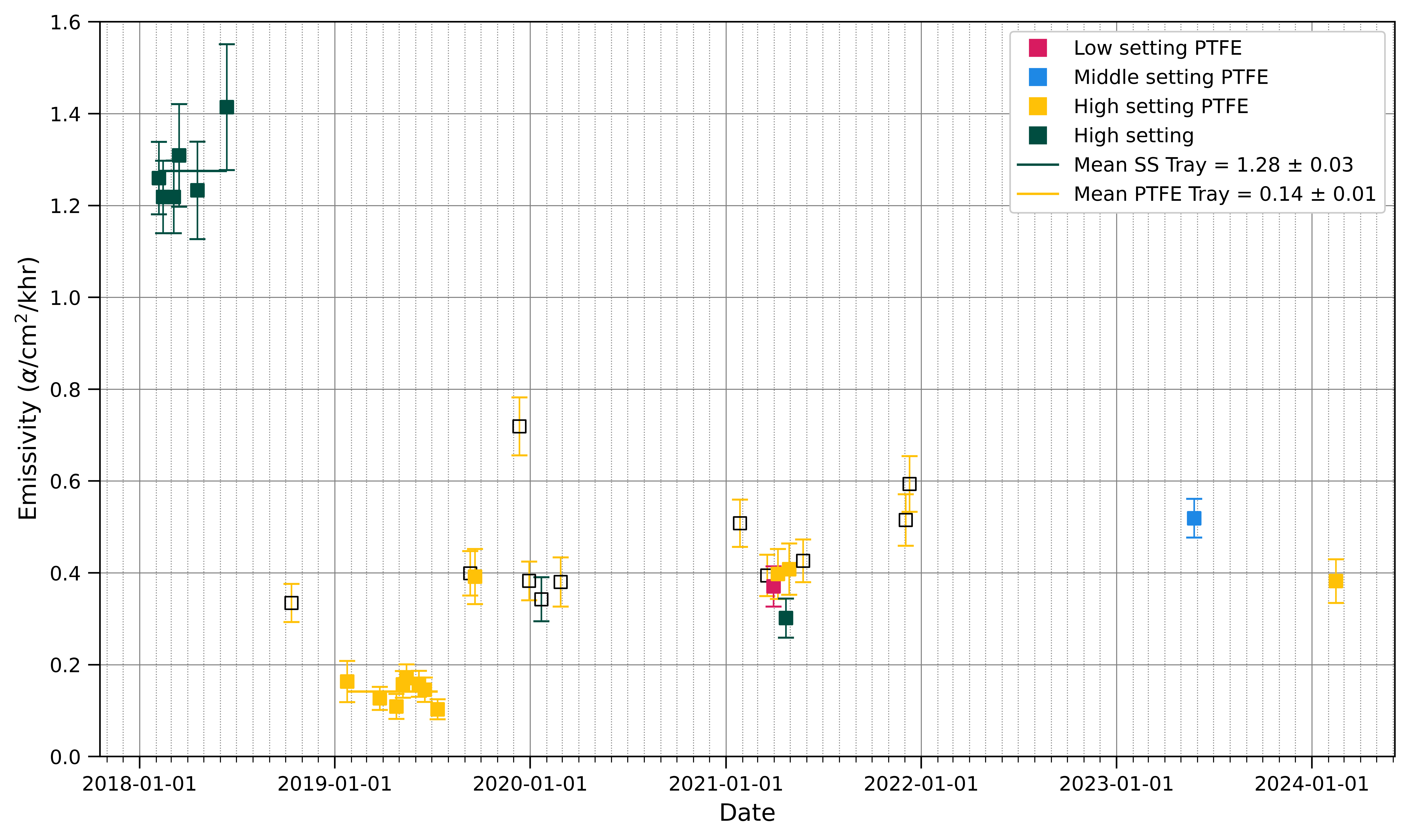}
\caption{Long-term stability study of the UltraLo-1800 event rate with different tray configurations. The filled green data points show the initial operation with the stainless steel sample tray. The filled yellow squares represent background measurements taken after a 48-hour purge; early data points (before the start of 2019) may reflect the system stabilizing after opening. The unfilled yellow squares represent measurements made immediately following sample assays, indicating the necessity of a purge period to return to the low background configuration. The study affirms the operational stability of the UltraLo-1800 and the effectiveness of the PTFE liner in reducing background interference.\label{fig:b}}
\end{figure}

The detector used for this study has been fully operational since January 2018. The sample tray is a critical component of the detector, as it holds the materials being assayed and determines the sensitive area exposed to the detector. To understand the inherent background measurements, we started by conducting measurements with the manufacturer-supplied stainless steel sample tray empty. A photograph of the stainless steel sample tray is shown in the left-hand panel of Figure~\ref{photo}. Several measurements were made, shown as the filled red data points in Figure~\ref{fig:b}, indicating an average rate of 1.28~$\pm$~0.03~$\alpha$/cm$^{2}$/khr. This result is in good agreement with results from other UltraLo-1800 facilities~\cite{nakib2013, McNally2014, xmass2018}. 

However, measurements of the event rate obtained with the sample tray empty only represent the background contribution from the tray itself. This does not accurately reflect the total background rate of the UltraLo-1800 during a sample measurement, as alpha emissions from other sources, such as the argon gas, can also contribute to the background signal. The reason lies in the properties of the samples and alpha particles. When a sample is inserted into the tray, the short range of alpha particles means any emissions from the sample are effectively prevented from being detected. Also, to meet the ideal measurement conditions, the sample should be designed to conform to the exact shape and size of one of the two standard measurement configurations. These are a circle with a diameter of 30~cm (area 700~cm$^{2}$) or a 43~cm $\times$ 43~cm rectangle (1800~cm$^{2}$). It is worth noting that for many samples, it is not possible to adhere to this requirement, leading to the activity from the tray, which is dependent on sample geometry, becoming dominant.

\begin{figure}[htbp]
\centering
\includegraphics[width=.7\textwidth]{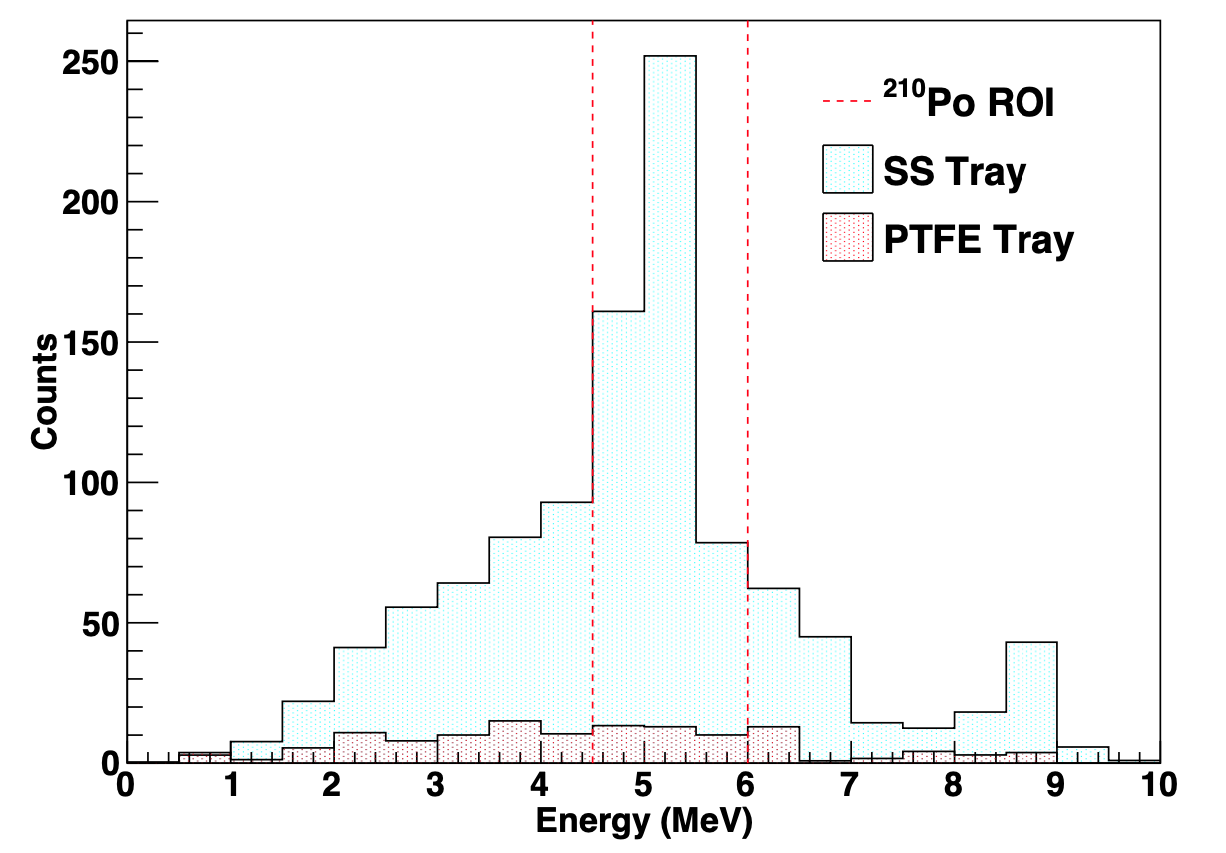}
\caption{Example UltraLo-1800 energy spectra resulting from 500-hour duration measurements with the manufacturer-provided stainless steel tray (blue histogram) and a tray formed from a 0.05~mm thick sheet of conductive-graphite filled PTFE (red histogram). In both cases, the UltraLo-1800 was otherwise empty.\label{fig:background_spectra}}
\end{figure}

One way to reduce the sample tray background is to insert an extremely radiopure liner into the tray, covering the detector's entire sensitive area. Here, we used a 0.05~mm thick sheet of conductive-graphite filled PTFE (DW 105 by DeWAL)~\cite{PTFEliner}, a photograph of which is shown in the right-hand panel of Figure~\ref{photo}. The impact of this liner on the background is presented in Figure~\ref{fig:background_spectra}, where the event yield as a function of energy for an exposure period of 500~hours is shown for the stainless steel tray, both with and without PTFE liner. With the liner, the event yield is significantly reduced.

Our long-term study of the UltraLo-1800 with this PTFE liner inserted (but otherwise empty) is presented as the filled yellow squares in Figure~\ref{fig:b}. The mean emissivity is measured to be 0.14~$\pm$~0.01~$\alpha$/cm$^{2}$/khr; individual measurements have a standard deviation of 0.01, indicating the stability of these measurements. It should be noted that some early background measurements, particularly those taken at the beginning of 2018, exhibit higher values than those observed later in the study. This is likely due to the system stabilizing after being opened to the atmosphere.

It is also shown in Figure~\ref{fig:b} as the unfilled yellow squares are measurements made with an identical detector configuration performed immediately following runs in which a sample had been assayed. As we can see, following a sample measurement, a significant purge period (typically 48 hours) is necessary to return the instrument to its lowest background configuration. 

Different research groups have adopted various purge periods for the UltraLo-1800, ranging from 1 hour \cite{ange2023characterization} to a full day \cite{xmass2018}. At Boulby, we have implemented a 48-hour purge period following sample assays. This extended period allows for the thorough removal of potential contaminants and ensures that the detector returns to a stable background level before subsequent measurements. The consistent stability observed in our long-term background measurements supports the effectiveness of this approach. While a systematic study of purge times would provide further insights into optimizing this procedure, it is beyond the scope of the current work and could be explored in future studies.



Further efforts to reduce background were made by replacing the tray with electroformed copper from the Pacific Northwest National Laboratory (PNNL). For transport from PNNL to Boulby, this copper sheet had been placed within three layers of heat-sealed 50 $\mu$m thick metallized mylar bags back-filled with nitrogen at atmospheric pressure. This packaging prevents exposure to radon and thus build up of surface plate-out. At Boulby, the electroformed copper was placed at the centre of the UltraLo-1800 and measured in `wafer mode', in which the detector's sensitive area is limited to that of the sample. An emissivity of 0.13~$\pm$~0.04~$\alpha$/cm$^{2}$/khr was recorded.

Similar efforts to quantify the absolute detector background have been reported by IBM~\cite{IBM} and XMASS~\cite{LRT_xmass}, resulting in 0.6~$\pm$~0.1~$\alpha$/cm$^{2}$/khr and 0.14~$\pm$~0.03~$\alpha$/cm$^{2}$/khr, respectively. The detector's operation near sea level may influence the IBM result, where cosmic radiation events contribute to the observed event rate. The XMASS detector, located underground at the Kamioka laboratory, produced results that are in statistical agreement with those reported here. The residual event rate observed in both the Boulby and Kamioka data is likely due to a combination of alpha emissions from detector construction materials and radon present in the argon buffer gas. Despite our best efforts, some background levels are irreducible due to the detector's material properties and operational requirements. We are exploring additional measures, such as argon scrubbing, to reduce this background further.

\section{Surface and Bulk Alpha Measurements}

Materials exposed to environments containing radon develop plate-out on their surfaces~\cite{nazaroff1988radon}. The dominant source of alpha-particle emission from high-purity material surfaces in environments with radon exposure is $^{210}$Po, due to its accumulation from radon progeny over the material's lifetime. The other decay progeny, a radionuclide with a significant half-life, $^{210}$Pb, decays almost entirely by beta particle emission~\cite{firestone19978th}. Consequently, unless exceptional care has been taken, most detector components manufactured in the last few decades will contain $^{210}$Po mixed into the bulk during production and manufacturing. By "bulk", we mean the internal substance of the material, which in some cases may include high surface area bulk products like powders for PTFE.

Alpha particle counting for samples often shows two primary components: 5.304~MeV alpha particles emitted by surface-deposited $^{210}$Po and lower energy alpha particles emitted from $^{210}$Po embedded within the bulk \cite{ZUZEL2017165}. The latter has lost some of its energy traversing to the material surface. Depending on the material, alpha particles from the bulk can be detected from various depths, up to several tens of micrometres in solids.

In our simulations using the GEANT4 framework, we modelled the copper sample with $^{210}$Po uniformly distributed throughout the bulk up to a depth of 12~micrometres from the surface. This depth corresponds to the maximum range of 5.3 MeV alpha particles in copper, beyond which alpha particles emitted in bulk would not have sufficient energy to escape the material and be detected by the UltraLo-1800 detector. The activity concentration of $^{210}$Po was assumed constant within this depth, reflecting uniform bulk contamination.

Each depth layer contributes to the overall spectrum based on its volume and the uniform activity concentration. The simulation accounts for energy losses of alpha particles as they traverse the copper matrix, leading to the continuous energy distribution observed for bulk emissions.

\begin{figure}[htbp]
\centering
\includegraphics[width=.8\textwidth]{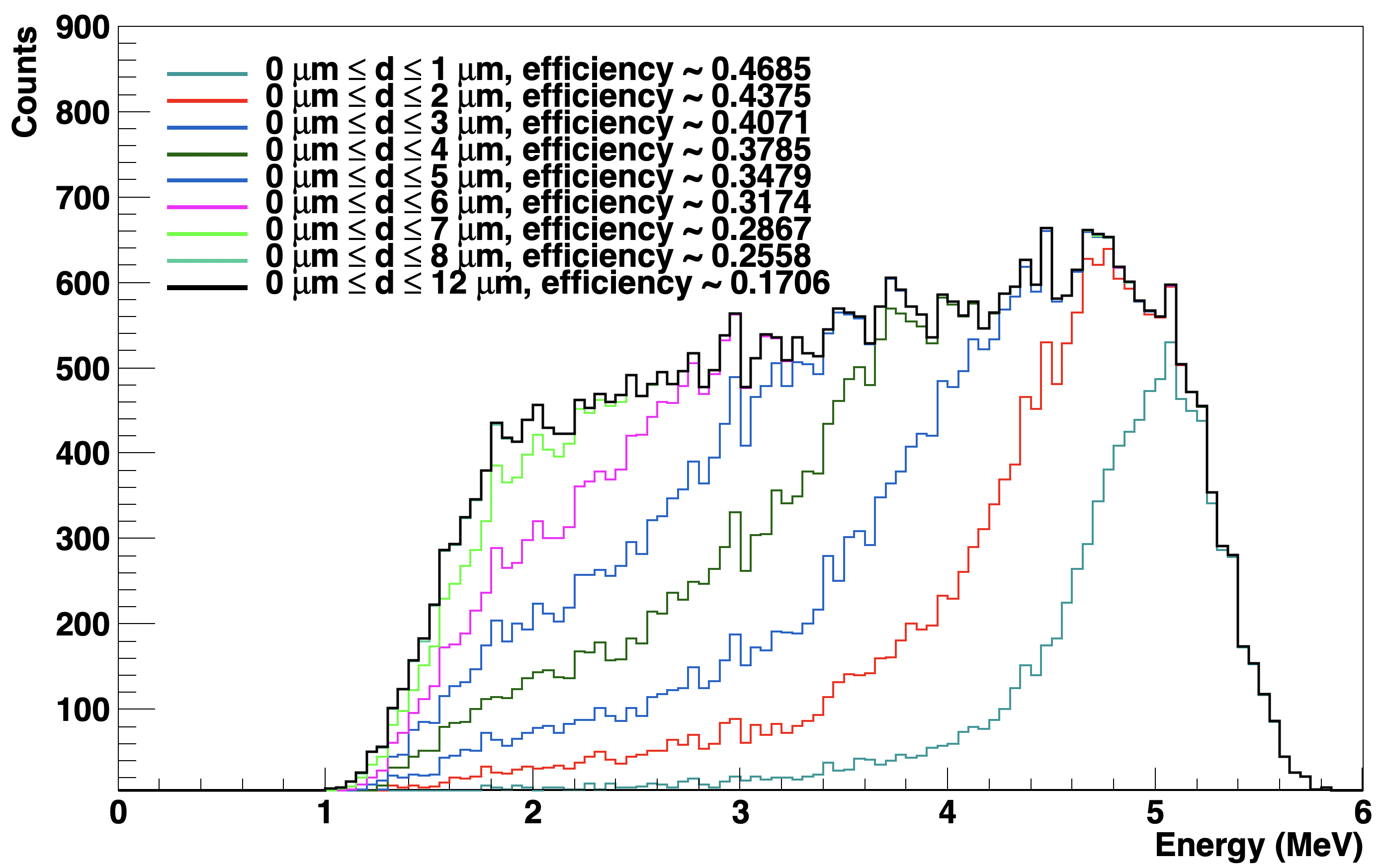}
\qquad
\includegraphics[width=.8\textwidth]{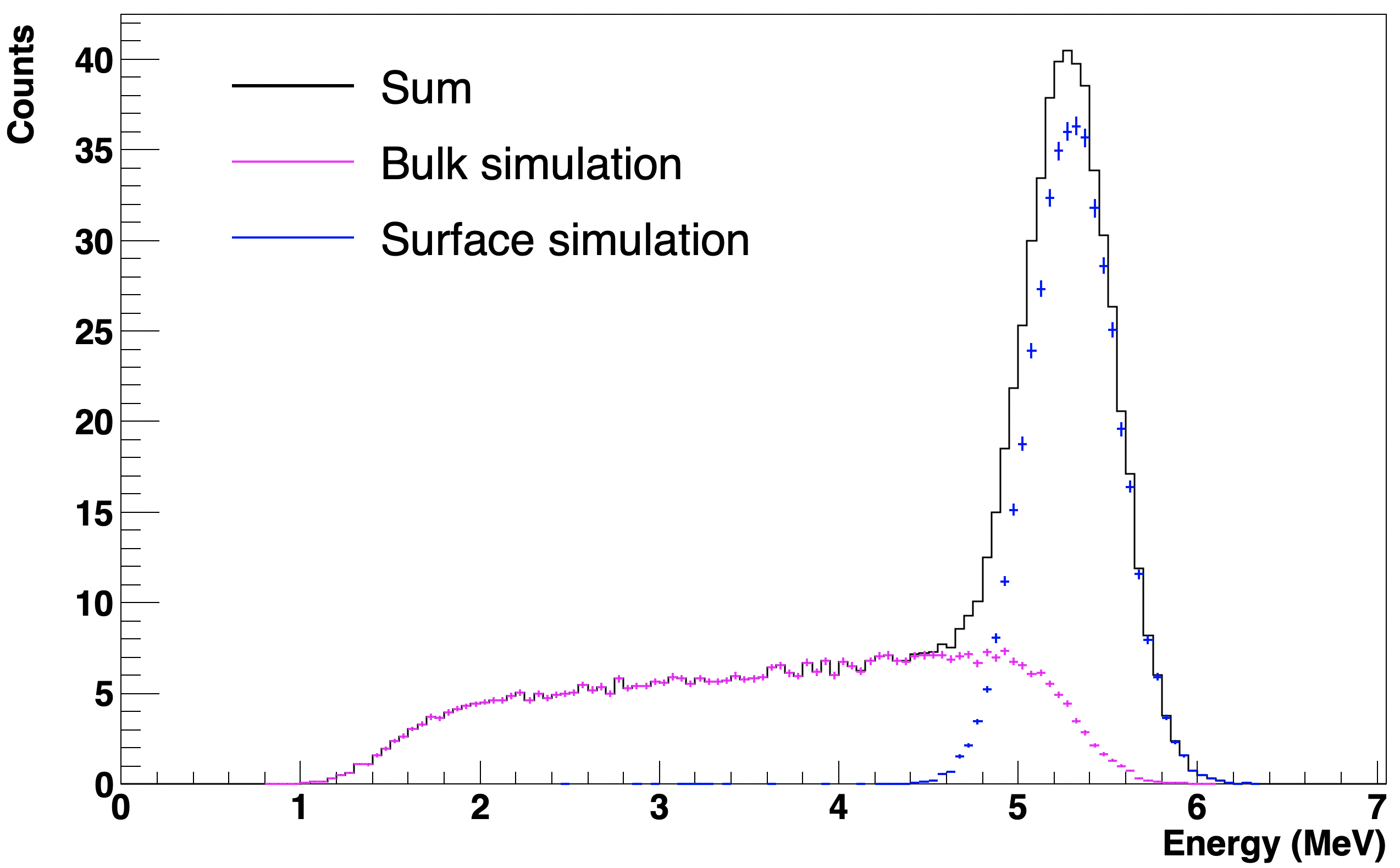}
\caption{Spectra illustrating simulations of energy depositions from surface and bulk $^{210}$Po alpha particles in copper. (Top) Detected alpha particle energy spectra from simulations of 10$^{5}$ $^{210}$Po decays in successive 1~$\mu$m thick slices of copper. (Bottom) Simulated combined surface and bulk $^{210}$Po alpha particle energy spectrum in copper.\label{fig:tibulk}}
\end{figure}

The GEANT4 framework was used to explore this behaviour, simulating alpha particles emitted from the surface and the bulk of a copper sheet covering the whole active area of the detector. The results are shown in Figure~\ref{fig:tibulk}. These results are generally consistent with those reported by XMASS \cite{xmass2018}. While their study provides valuable insights into alpha particle behaviour, it is important to note that their simulation methodology, as described in their publication, did not appear to incorporate a threshold value. This difference in approach may lead to variations in the interpretation and application of the results, particularly when considering the practical limitations of the detector threshold.

The left-hand panel of Figure~\ref{fig:tibulk} predicts the alpha particle yield detected from emission throughout layers of sequentially increasing depth, illustrating how emission from different depths contributes to the overall rate. Regarding bulk emission, the efficiencies noted in the left-hand figure illustrate how decay depth impacts the probability of an alpha particle escaping the bulk. Decays occurring near the surface have almost an equal chance of being emitted or buried within the sample. However, decays occurring at greater depths have progressively smaller chances of escaping the material.

The right-hand panel displays the total bulk emission from this simulation, combined with a simulation of $^{210}$Po decaying on the copper sheet's surface. It should be noted that copper serves as a model substance here, and the simulation results would show similar features for other materials used in detector fabrication.

During sample measurements, additional alpha particle signals are expected due to the decay of radioisotopes present in the detector's argon gas. This would generate broadly Gaussian features at 6.115~MeV (from decays of $^{218}$Po), 6.207~MeV ($^{212}$Bi),  6.906~MeV ($^{216}$Po), 7.834~MeV ($^{214}$Po) and 8.954~MeV ($^{212}$Po)~\cite{wang2021ame}, note that these values represent the total decay energies (Q-values) for each decay, and the alpha particles emitted carry slightly less energy due to the recoil of the daughter nucleus. The detected yields for each of these contributions reflect the activity level of the radionuclide in the gas, its decay branching ratio, and the detector efficiency at that energy~\cite{XIA_usersmanual}. Absolute detector efficiencies were based on data from XIA, validated at 5.304~MeV using $^{210}$Po calibration data, and extrapolated as required to other energies.

\subsection{Assay of Copper Plate-out and Bulk Activity}

In line with future requirements for rare event search experiments, we performed measurements of surface and bulk alpha particle emanation from copper sheeting to optimise cleanliness procedures. An industrial-grade copper sheet was assessed before and after prolonged exposure to environmental radon, enabling the accumulation of $^{210}$Po. 

The copper sheet, measuring 43~cm $\times$ 43~cm and with a thickness of 0.6~mm, was positioned at the centre of the PTFE sample tray for measurements. These dimensions were chosen to maximise detection efficiency and minimise background contributions from the sample tray.

\begin{figure}[htbp]
\centering
\includegraphics[width=.8\textwidth]{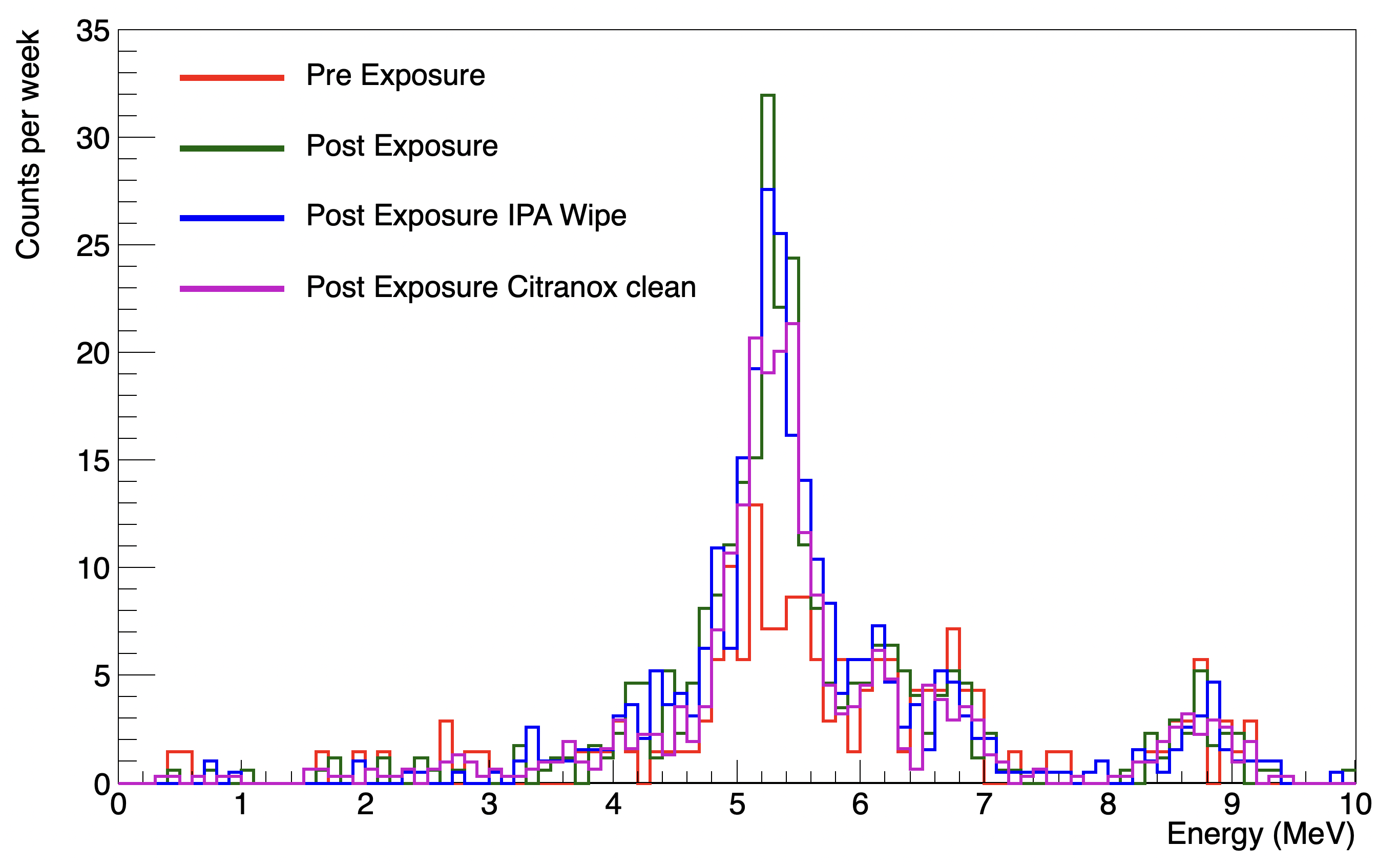}
\caption{Spectra of energy depositions from alpha particles originating from radon plate-out on a copper sample: prior to exposure to radon-contaminated air within the laboratory (red); after 458 days of exposure within the laboratory (green); after cleaning with an IPA wipe (blue); and following cleaning with Citranox (purple).\label{fig:copper}}
\end{figure}

Initially, the copper sheet was thoroughly cleaned with Isopropyl Alcohol (IPA) and allowed to air dry before being installed in the detector. Although we know that IPA is ineffective in removing all dust or contamination, it was used as a starting point for our cleaning procedure. The results produced an activity of 0.247 $\pm$ 0.024 mBq for $^{210}$Po, as illustrated in Figure~\ref{fig:copper}.

The copper sheet was then placed in the BUGS cleanroom, where it was exposed to 2.30 $\pm$ 0.03~Bq/m$^{3}$ of radon for 458 days. This exposure duration was chosen based on our prior experience with radon exposure and aimed to achieve significant plate-out of radon progeny, including $^{210}$Po, onto the copper surface, thus increasing the activity levels.

Subsequently, the sample was reinserted into the detector without additional cleaning and measured for 15 days. The detected activity in the region of interest for $^{210}$Po was then 0.504 $\pm$ 0.022 mBq, indicating a significant increase in activity.

Next, the sample was cleaned using IPA again and air-dried. This resulted in a measured activity of 0.485 $\pm$ 0.020 mBq in the same regions of interest. Finally, the copper was cleaned using CitraNOX~\cite{citranox2023}, which resulted in a measured activity of 0.336 $\pm$ 0.013~mBq for $^{210}$Po. A comparison of the copper's $^{210}$Po activities following the two cleaning methods indicates that the CitraNOX acid was more effective in reducing $^{210}$Po contamination from the copper surface compared to cleaning with IPA. This observation is consistent with findings from previous research, which demonstrated CitraNOX's superior efficacy over IPA~\cite{scovell2023radioassay, stein2018radon}. However, it should be mentioned that these cleaning methods primarily affect surface contamination and would not be expected at affect bulk activity significantly.

Potential reasons for the remaining alpha activity post-CitraNOX cleaning include the presence of embedded $^{210}$Po within the copper's bulk, which is not removed by surface cleaning. Also, the copper's microstructures or porosities may trap contaminants, making them less vulnerable to cleaning agents.

To further optimize the cleaning process, future investigations will focus on a combination of mechanical and chemical cleaning techniques. This will include exploring the effectiveness of ultrasonic cleaning combined with chemical agents, potentially targeting embedded contaminants more effectively. Additionally, systematic studies should be conducted to rigorously evaluate the efficacy of different cleaning protocols on various materials commonly used in detector construction.

Overall, our cleaning methods successfully reduced alpha particle emissions from the copper surface. However, even after using CitraNOX acid wipes, which we have previously found to be more effective than IPA in reducing alpha activity \cite{scovell2023radioassay}, the alpha activity was still above the desired level for sensitive detectors. Therefore, further optimisation of the cleaning process is required, although our study has shown a clear pathway to follow for improvements (see Table \ref{tab:0vBBbackgrounds}).

\begin{table}[htbp]
    \centering
    \caption{The measured activities of $^{210}$Po emanating from a copper sheet along with implications of each state for cleanliness procedures.}
    \label{tab:0vBBbackgrounds}
    \begin{tabularx}{\textwidth}{ 
        >{\raggedright\arraybackslash}X 
        >{\raggedright\arraybackslash}X 
        >{\centering\arraybackslash}X 
        >{\raggedright\arraybackslash}X }
        \toprule
        \textbf{Name} & \textbf{Description} & \textbf{$^{210}$Po Surface Activity ($\mu$Bq/cm$^{2}$)} & \textbf{Implications} \\
        \midrule
        Pre-exposure & Assayed when the copper sheet first arrived at the laboratory, without having any exposure to air. & $0.137 \pm 0.013$ & Baseline measurement for comparison with subsequent measurements. \\
        Post-exposure & Assayed after the copper sheet had been left inside the laboratory for 458 days. & $0.280 \pm 0.012$ & Demonstrates increase in activity due to radon exposure. \\
        Post-exposure IPA wipe & Assayed after wiping the post-exposure copper sheet using IPA. & $0.269 \pm 0.011$ & Indicates IPA's effectiveness in reducing activity, though modest. \\
        Post-exposure Citranox clean & Assayed after wiping the post-exposure copper using IPA, then Citranox acid. & $0.187 \pm 0.007$ & Reveals significant reduction in activity, highlighting Citranox's superior cleaning efficiency. \\
        \bottomrule
    \end{tabularx}
\end{table}

\section{Conclusion}

The XIA UltraLo-1800 detector, installed within the Boulby UnderGround Screening (BUGS) facility, has been characterised and shown to maintain stable internal backgrounds over extended periods. This stability is essential for detecting minute levels of radioactivity in materials used for rare-event searches. A GEANT4 simulation model was developed to effectively differentiate between bulk and surface alpha emissions, a distinction critical for understanding and mitigating contamination sources in detector materials. Specifically, our model simulates the behaviour of alpha particles emitted from both the surface and bulk, providing insights into how emission depth affects detection probabilities.

Based on the study of background rates from a variety of trays used in the XIA, we concluded that the PTFE-lined tray reduced the background rate by a factor of 9 compared to the default XIA stainless steel (SS) tray. However, further measurements showed that the ultra-pure electroformed copper tray produced the lowest background rate, and therefore the plan is to switch to the electroformed copper tray for future measurements. The reduced background rate is also directly attributable to operating within a deep underground laboratory, which effectively shields the detector from cosmic ray-induced backgrounds. Moreover, the UltraLo-1800 detector has been employed to quantify the effectiveness of various surface-cleaning techniques for materials commonly used in particle physics, such as copper. Through this work, we have identified more effective methods for reducing surface contamination, particularly crucial for mitigating radon daughter plate-out.

In this work, we have demonstrated the UltraLo-1800 detector's capability to measure both surface and bulk alpha emissions, aided by our GEANT4 simulation model. This capability establishes the UltraLo-1800 as an essential component of the BUGS facility's material screening program. These results have significant implications for future rare-event search experiments, such as BUTTON (Boulby Underground Technology Testbed for Observing Neutrinos) \cite{kneale2023button} and XLZD \cite{baudis2024darwin}. The advancements presented in this study pave the way for achieving the required levels of material purity and background suppression in these next-generation experiments.

Ongoing efforts to enhance BUGS’s material screening capabilities include the introduction of inductively coupled plasma mass spectrometry (ICP-MS) and a radon emanation detection system. Additionally, plans are underway to integrate in-house production of electroformed copper, which will further reduce background contamination and improve the sensitivity of future assays. These enhancements will solidify BUGS as a leading facility in underground science and rare-event detection, supporting the stringent demands of upcoming experiments.




\bibliographystyle{JHEP}
\bibliography{main.bib}


\end{document}